\documentclass[prb,twocolumn,showpacs,floatfix,superscriptaddress]{revtex4}
\usepackage{amssymb}

\usepackage{graphicx}
\usepackage{dcolumn}
\usepackage{bm}
\usepackage{latexsym}
\usepackage{amsmath}

\begin{document}

\title{Spin and charge ordering in three-leg ladders in oxyborates.}
\author{E. Vallejo and M.\ Avignon}
\affiliation{Laboratoire d'Etudes des Propri\'{e}t\'{e}s Electroniques des Solides
(LEPES)-Centre National de la Recherche Scientifique (CNRS), BP\ 166, F-38042\ Grenoble Cedex 9, France.}
\date{\today}

\begin{abstract}
We study the spin ordering within the 3-leg ladders present in the 
oxyborate
Fe$_{3}$O$_{2}$BO$_{3}$ consisting of localized classical spins 
interacting
with conduction electrons (one electron per rung).\ We also consider 
the
competition with antiferromagnetic superexchange interactions to 
determine
the magnetic phase diagram. Beside a ferromagnetic phase we find (i) a 
phase
with ferromagnetic rungs ordered antiferromagnetically (ii) a zig-zag 
canted
spin ordering along the legs. We also determine the induced charge 
ordering
within the different phases and the interplay with lattice instability. 
Our
model is discussed in connection with the lattice dimerization 
transition
observed in this system, emphasizing on the role of the magnetic 
structure.
\end{abstract}

\maketitle

\affiliation{Laboratoire d'Etudes des Propri\'{e}t\'{e}s Electroniques 
des Solides
(LEPES)-Centre National de la Recherche Scientifique (CNRS), BP\ 166, 
F-38042\ Grenoble Cedex 9, France.}

The ordering of the local spins interacting with conduction electrons
remains an important problem and has become very active in the context 
of
manganites. The coupling can be antiferromagnetic as in heavy-fermions
systems or Kondo insulators or ferromagnetic as a result of Hund's 
coupling
in manganites. This gives rise to the general double exchange (DE)
interactions\cite{anderson} favoring a ferromagnetic background of 
local
spins. This ferromagnetic tendency is expected to be thwarted by
antiferromagnetic superexchange (SE) interactions between the localized
spins leading to interesting and unusual magnetic states. Instead of 
the
canted states conjectured by de Gennes\cite{degennes}, spin ordering
consisting of ferromagnetic \textit{islands} coupled 
antiferromagnetically
has been identified for various commensurate fillings both for $S=1/2$
quantum spins in one dimension\cite{garcia00} and classical spins in 
two
dimensions\cite{aliaga01}. Carriers are found to be localized in the
ferromagnetic islands giving rise to bond ordering and as a consequence
leads to charge ordering.

The ludwigite oxy-borate system Fe$_{3}$O$_{2}$BO$_{3}$ may provide 
evidence
of this mechanism for the existence of simultaneous spin and charge 
ordering
resulting from the competition between DE and SE. Fe-ludwigite contains
subunits in the form of 3-leg ladders (3LL) of Fe cations and presents 
an
interesting structural and charge ordering transition at $T_{c}\approx 
283K$%
, such that long and short bonds on the rungs alternate along the 
ladder
axis \cite{mir01}. As evidenced by M\"{o}ssbauer studies\cite
{larrea01,larrea04} and X-ray diffraction\cite{bordet} each rung can be
viewed as three Fe$^{3+}$ ions (\textit{triad}) with high-spin $S=5/2$ 
local
spins sharing an extra itinerant electron. The charge distribution 
among the
triads is a key issue. Spin ladders have recently attracted a 
considerable
interest but we have here an interesting case of a spin ladder coupled 
with
conduction electrons. The coupling is similar to the one encountered in 
Fe
double-perovskite systems\cite{carvajal05}. In the Fe$^{3+}$ $d^{5}$
configuration all orbitals being occupied in one spin channel, 
itinerant
electrons can hop to a site $i$ only if its spin is antiparallel to the
local spin $\overrightarrow{S_{i}}$. This is indeed equivalent to DE 
with an
effective antiferromagnetic and \textit{infinite exchange integral}.
Antiferromagnetic SE interactions resulting from virtual hopping among 
the
Fe-$d^{5\text{ }}$ configurations have been estimated, leading to 
strongly
interacting spin units of the Fe-triads in which all nearest-neighbor 
(n.n)
spins are antiferromagnetically coupled both above and below the 
structural
transiton temperature\cite{whangbo02} in contradiction with recent 
neutron
results\cite{bordet}. In addition, it can be shown that an homogeneous
magnetic phase is not compatible with the observed charge distribution 
on
the different Fe sites. In this letter we will show that the inclusion 
of
the interaction between intinerant electrons and local spins will
drastically improve this picture.

Since the local spins $\overrightarrow{S_{i}}$ are fairly large $S=5/2$ we
will treat them as classical spins specified by their polar angles $\theta
_{i}$ and $\varphi _{i}$ ($0<\theta _{i}<\pi ,0<\varphi _{i}<2\pi $) defined
as usual with respect to a $z$-axis taken as the spin quantization axis of
itinerant electrons. Rotating the itinerant electron quantization axis on
each site to make it parallel to $\overrightarrow{S_{i}}$, one gets the
rotated electron operators with spin opposite to the local spin $c_{i}^{+}$($%
c_{i}$) in terms of the original electron operators $d_{i\sigma }^{+}$($%
d_{i\sigma }$) as $c_{i}^{+}=\cos (\theta _{i}/2)d_{i\downarrow
}^{+}-e^{-i\varphi _{i}}\sin (\theta _{i}/2)d_{i\uparrow }^{+}$. The rotated
electrons are indeed \textit{spinless }electrons. The effective hopping
between these electrons antiparallel to local spins at sites $i$ and $j$ is
therefore given by $t_{i,j}^{e}=t_{\nu }(\cos \frac{\theta _{i}}{2}\cos 
\frac{\theta _{j}}{2}+e^{-i(\varphi _{i}-\varphi _{j})}\sin \frac{\theta _{i}%
}{2}\sin \frac{\theta _{j}}{2})$, $t_{\nu }=t_{a},t_{c}$ being the nearest
neighbor (n.n) hopping integrals on the rungs and along the axis of the
ladder.

So, to describe the magnetic structure, we represent the interaction between
the Fe$^{3+}$ localized spins $\overrightarrow{S_{i}}$ and the itinerant
electrons by the tight-binding Hamiltonian together with SE interactions
among the local spins

\begin{equation*}
H=-\sum_{\left\langle ij\right\rangle }(t_{i,j}^{e}c_{i}^{+}c_{j}+\mathit{%
h.c.})+\sum_{\left\langle ij\right\rangle }J_{ij}\overrightarrow{S_{i}}\cdot 
\overrightarrow{S_{j}}
\end{equation*}

\begin{figure}[t]
\centering
\includegraphics[scale=0.3]{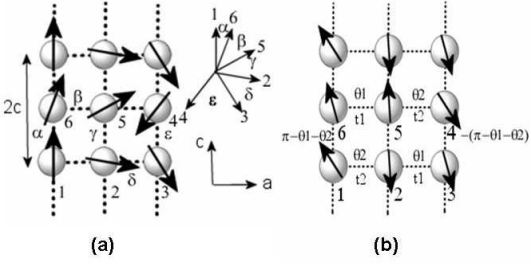}
\caption{(a) Magnetic structure of the 3-leg ladder Ludwigite. The five
angles $\protect\alpha ,\protect\beta ,\protect\gamma ,\protect\delta 
,%
\protect\epsilon $ gives the orientation of the spins on the six sites 
$%
i=1-6 $ unit cell. (b) Magnetic structure of the $I_{a}$ phase. This
structure can present a zigzag modulation of the angles $\protect\theta 
_{1}$
and $\protect\theta _{2}.$}
\label{Figure1}
\end{figure}

$\left\langle ij\right\rangle $ represents n.n sites.\ We further assume
that this band is non-degenerate, therefore the band filling is $n=1/3$. We take the simple
situation in which all the spins are in the same plane. This simplification
is inspired by Monte Carlo simulations on 2D systems in which noncoplanar
spin configurations never seem to appear\cite{aliaga01} and is consistent
with neutron scattering results \cite{bordet}. All coplanar phases being
degenerate, we choose the plane of the ladder, taking $\theta _{i}=\pi /2$
and the hopping terms simply becomes $t_{i,j}^{e}=\frac{t_{\nu }}{2}%
(1+e^{-i(\varphi _{i}-\varphi _{j})}).\;$Guided by the periodicity $2c$ of
the low temperature distorted phase\cite{mir01}, we consider a unit-cell
containing two rungs. We define the magnetic structure by the five angles $%
\alpha ,\beta ,\gamma ,\delta ,\varepsilon $ giving the orientation of the
spins on the six sites $i=1-6$ unit cell as shown in Fig. \ref{Figure1}-a. $%
J_{ij}=J_{a},J_{c}$ are SE interactions in the two directions.

\begin{figure}[t]
\centering
\includegraphics[scale=0.55]{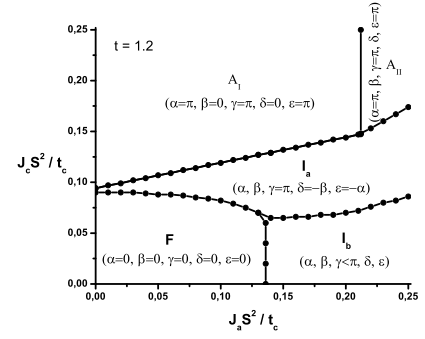}
\caption{Phase diagram as function of $\frac{JaS^{2}}{t_{c}}$ and 
$\frac{%
JcS^{2}}{t_{c}},$ for a typical value of $t=\frac{t_{a}}{t_{c}}=1.2.$ 
The
different phases are described in the text.}
\label{Figure2}
\end{figure}

The kinetic energy term favors a ferromagnetic arrangement of the local
spins which competes with the SE, leading to a variety of complex
structures. After the Fourier transformation, the dispersion of the
conduction electrons is obtained from the tight-binding matrix with the
wave-vector $k$ in the $c$-direction, $-\pi /2c<k<\pi /2c$. In the 
general
case, it consists of six bands $\epsilon _{i=1-6}(k)$, the values 
$\epsilon
_{i}(k)$ are increasing from $i=1$ to $6$. For the band-filling $n=1/3$ 
the
two lowest bands $\epsilon _{1,2}$ only are occupied. We minimize the 
total
energy with respect to the five angles $\{\alpha ,\beta ,\gamma ,\delta
,\varepsilon \}$. Fig. \ref{Figure2} shows the phase diagram as 
function of $%
J_{a}S^{2}/t_{c}$ and $J_{c}S^{2}/t_{c}$ for a typical value 
$t=\frac{t_{a}}{%
t_{c}}=1.2$ roughly estimated from the different Fe-Fe distances in the
triad and along the legs. 
Besides the fully ferromagnetic ($F$) state characterized by the 
uniform
angles $\alpha =\beta =\gamma =\delta =\varepsilon =0$, when $J_{a}$ 
and $%
J_{c}$ are not too large $J_{c}S^{2}/t_{c}\lesssim 0.07$ and $%
J_{a}S^{2}/t_{c}\lesssim 0.13$, we find two other phases (i) at larger 
$%
J_{c} $ a phase $A$ which is antiferromagnetic in the $c$-direction 
($\alpha
=\gamma =\varepsilon =\pi $) with two different angles in the rung (ii) 
a
phase $I$ with different angles ($\alpha ,\beta ,\gamma ,\delta 
,\varepsilon 
$) which is the stable one in a large part of the phase diagram for 
lower $%
J_{c}$. These phases are further described below.\ Except for the
ferromagnetic phase, there is a gap between the two lowest bands and 
the
middle ones.\ At $1/3$-filling the Fermi energy is located in this gap 
so
that all these phases are insulating. For symmetry reason the same 
occurs
also at $2/3$-filling. This gap depends on the values of the different
angles and can be direct or indirect.

In phase $A$ the hopping is totally suppressed in the $c$-direction and 
the
dipersion reduces to three energy levels. The particular phase $A_{I}$ 
with
fully ferromagnetic rungs ($\beta =\delta =0)$ is encountered at lower 
$%
J_{a} $. This phase is in qualitative agreement with the magnetic 
structure
recently proposed from neutron experiments at $82K$\cite{bordet}.\ It 
is
indeed very similar to a phase already found with Monte-Carlo 
calculations
in the 2D model\cite{aliaga01}.\ At larger $J_{a}$, canting occurs 
within
the rungs with two different angles $\beta $, $\delta $, we call this 
phase $%
A_{II}$.

Phase $I$ presents very interesting simple structures as, for example, 
the
phase $I_{a}$ ($\alpha ,\beta ,\gamma =\pi ,\delta =-\beta ,\varepsilon
=-\alpha $) which can be defined in terms of only two angles $\theta 
_{1}$%
and $\theta _{2}$ ($\beta =\theta _{1},\alpha =\pi -\theta _{1}-\theta 
_{2}$%
).\ It is AF along the central leg so that no hopping is taking place 
along
this leg. This structure presents a zig-zag modulation of the angles 
$\theta
_{1}$and $\theta _{2}$ and, consequently of the hopping $t_{1\text{,}}$ 
$%
t_{2}$ as shown in Fig. \ref{Figure1}-b. A phase called $I_{b}$ tends 
to a
ferromagnetic behavior along $c$-direction with $\gamma <\pi $.

As soon as the central leg is AF ($\gamma =\pi $) the bands are 
two-fold
degenerate with gaps at $k=\pm \frac{\pi }{2c}$. The dispersion of the 
bands
are $\epsilon (k)=0$ and 
\begin{eqnarray*}
\epsilon (k) &=&\pm \lbrack 1+t^{2}+\frac{1}{2}t^{2}(\cos \theta 
_{1}+\cos
\theta _{2})-\cos (\theta _{1}+\theta _{2})+ \\
&&\left. +(1-\cos (\theta _{1}+\theta _{2}))\cos 2kc\right] 
^{\frac{1}{2}}.
\end{eqnarray*}
The lower band is filled, precisely for $n=1/3$, lowering the kinetic 
energy
to stabilize this phase. The total energy per rung $E$ can be expressed as 
\begin{eqnarray*}
\frac{E}{t_{c}} &=&-\frac{2}{\pi }[2+t^{2}+\frac{1}{2}t^{2}(\cos \theta
_{1}+\cos \theta _{2})+ \\
&&-2\cos (\theta _{1}+\theta _{2})]^{\frac{1}{2}}\mathcal{E(}q)-2J_{c}\frac{%
S^{2}}{t_{c}}[\frac{1}{2}+\cos (\theta _{1}+\theta _{2})] \\
&&+\frac{J_{a}S^{2}}{t_{c}}(\cos \theta _{1}+\cos \theta _{2}),
\end{eqnarray*}
$\mathcal{E(}q)$ being the complete Elliptic Integral of second kind 
with
parameter $q=\frac{2(1-\cos (\theta _{1}+\theta 
_{2}))}{2+t^{2}+\frac{1}{2}%
t^{2}(\cos \theta _{1}+\cos \theta _{2})-2\cos (\theta _{1}+\theta 
_{2})}$.

\begin{figure}[t]
\centering
\includegraphics[scale=0.58]{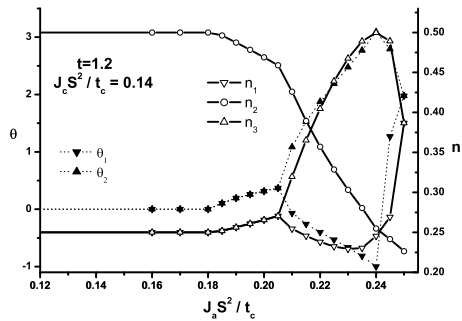}
\caption{Cut of the phase diagram along the line 
$\frac{JcS^{2}}{t_{c}}%
=0.14. $ showing the angles between the spins and the resulting 
charges.}
\label{Figure3}
\end{figure}

The angle $\gamma $ varies discontinuously between phase $I_{a}$ 
($\gamma
=\pi $) and phases $F$ ($\gamma =0$) and $I_{b}$ ($\gamma <\pi $), so 
these
transitions are first order. All other transitions are second order. In 
the $%
I_{b}$ phase, close to $F$ we find a canted ferromagnetic phase with 
canting
within the rungs, one angle only $\beta $ (or equivalently \ $\delta $)
being different from zero; at the transition $\beta \rightarrow 0$ 
giving
the second order boundary line $\frac{J_{a}S^{2}}{t_{c}}=\frac{t\arccos 
(-t/2%
\sqrt{2})}{4\pi \sqrt{2}}$. Between $F$ and $A_{I}$ phases, the $I_{a}$
phase has essentially $\theta _{1}=\theta _{2}=\theta $; this can be 
seen
for example, close to $A_{I}$, in Fig. \ref{Figure3} for 
$\frac{J_{c}S^{2}}{%
t_{c}}=0.14$.\ Therefore the transition line between $I_{a}$ and 
$A_{I}$ ($%
\theta _{1}$, $\theta _{2}\rightarrow 0$) is also second order 
corresponding
to $\frac{J_{c}S^{2}}{t_{c}}=\frac{\sqrt{2}(4-t^{2})}{32t}+$ $\frac{%
J_{a}S^{2}}{4t_{c}}$. For larger values of $\frac{J_{a}S^{2}}{t_{c}}$ 
the
phase evolves towards the more general zig-zag structure $\theta 
_{1}\neq
\theta _{2}$ (see Fig. \ref{Figure3}).

As we mentioned the charge distribution is crucial in the Fe-ludwigite
ladder so let us examine this point in detail. It is clear that bond
ordering is linked to the spin ordering through the modulation of the
hopping amplitudes.\ The ferromagnetic bonds tend to localize the extra
electron. This in turn may induce different types of charge ordering on 
the
non-equivalent Fe-sites in the rung. 
Experimentally\cite{larrea01,larrea04}
two charge regimes are identified (i) above $T_{c}$, the side sites 1 
and 3
are identical $n_{1}=n_{3}\sim 0.25-0.3$ while the central site 2 has 
more
electrons $n_{2}\sim 0.5$ (ii) below $T_{c}$ down to $74K$ the charge 
on
site 3 (the site which gets closer to site 2) increases close to the 
charge
of site 2 which remains stable, $n_{2}\approx n_{3}\sim 0.5$, and at 
the
same time the charge of site 1 decreases to $n_{1}\sim 0.15$. Of course
these values\cite{bordet} indicate only the tendencies, since one 
should
have $n_{1}+n_{2}+n_{3}=1$. However below $74K$ two contradictory 
behaviours
have been reported \cite{larrea04,douvalis02}. Douvalis et al\cite
{douvalis02} found that the low temperature ordering below $T_{c}$ 
persists
down to $T=0$, while Larrea et al\cite{larrea04} recover the same 
charge
ordering as above $T_{c}$.

\begin{figure}[t]
\centering
\includegraphics[scale=0.23]{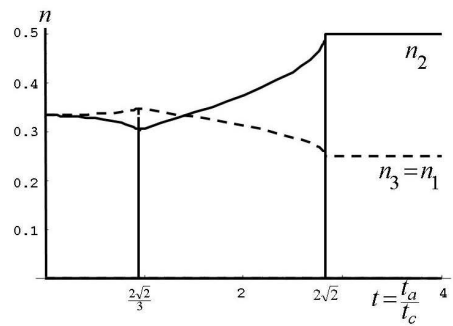}
\caption{Electronic distribution $n_{i}$ of the homogeneous magnetic 
phase
as function of $t.$ }
\label{Figure4}
\end{figure}

To begin with, let us look at homogeneous magnetic phases i.e a phases
without modulation of the hopping amplitude; in this sense 
ferromagnetic and
paramagnetic phases are equivalent, only the effective hoppings are
different in the two cases. The electronic distribution is shown in 
Fig. \ref
{Figure4} as a function of $t$. 
We see that the high temperature behaviour can be reproduced only if 
$t$ is
large $t\gtrsim 2.5-3$, in particular for $t\geq 2\sqrt{2}$ one gets $%
n_{1}=n_{3}=1/4$ and $n_{2}=1/2$, but such $t$ values are far too large 
in
the Fe-ludwigite ladder. But we see that the same regime can be reached 
in
the $A_{I}$ phase as well since, in this case, the effective hopping is 
zero
in the $c$-direction which is equivalent to taking $t_{c}=0$ (see Fig. 
\ref
{Figure3}) and the problem reduces to three sites. The $I_{a}$ phase 
close
to $A_{I}$ with $\theta _{1}=\theta _{2}$ could also give quite well 
the
high temperature charge distribution as seen in Fig. \ref{Figure3} for 
$%
J_{a}S^{2}/t_{c}\lesssim 0.2$. However, as can be seen in Fig. 
\ref{Figure3}%
, an interesting point resulting from our analysis is the existence of 
the$\
I_{a}$ structure with $\theta _{1}\neq \theta _{2}$ as in Fig. 
\ref{Figure1}%
-b. This produces a zig-zag bond alternation which, in turn, will give 
rise
to a lattice instability of the same type. Due to the magnetic 
structure the
two border sites of a rung have different electronic charges leading to 
the
formation of a zig-zag charge ordering, $n_{2}\approx n_{3}\gg n_{1}$,
similar to the one observed experimentally below $T_{c}$. Note that a 
phase
of type ($\theta _{1}=0,\theta _{2}=\pi /2$), $\uparrow \uparrow
\longrightarrow $ on the rung, has been proposed at $10K$\cite{bordet} 
in
contrast with the antiferromagnetic ordering $\uparrow \downarrow 
\uparrow $
inside the triad obtained from earlier neutron 
experiments\cite{attfield92}.
Except asymptotically i.e. $J_{a}\rightarrow \infty $, we do not find 
phases
with AF arrangement of the triads.

\begin{figure}[t]
\centering
\includegraphics[scale=0.6]{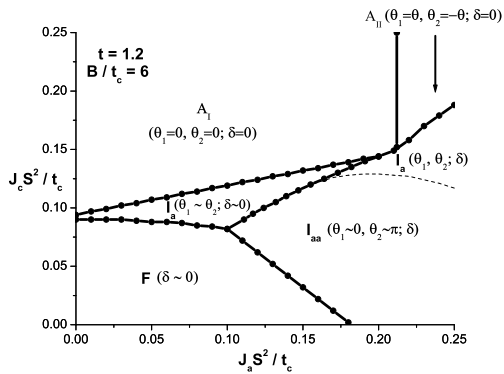}
\caption{$F-I_{a}$ phase diagram for the elastic parameter $%
B/t_{c}=6$ as a function of $%
\frac{JaS^{2}}{t_{c}}$ and $\frac{JcS^{2}}{t_{c}}.$ Note that $A_{I}$ and
$A_{II}$ are particular cases of $I_{a}$. The
distorted phase $I_{aa}$ occurs below the dashed line.}
\label{Figure5}
\end{figure}

\begin{figure}[t]
\centering
\includegraphics[scale=0.35]{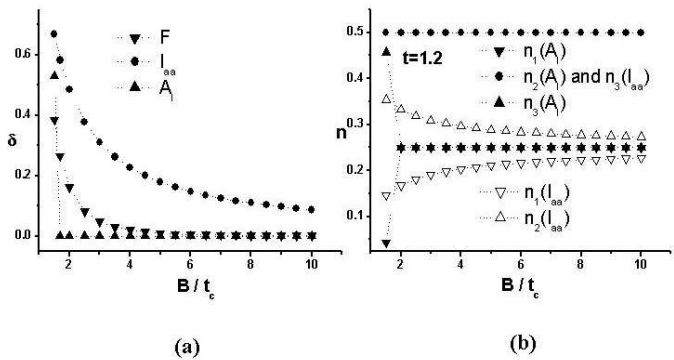}
\caption{(a) Lattice distortion of the rung among the $F$, $A_{I}$ and 
$%
I_{aa}$ phases as function of $B/t_{c}.$ (b) The corresponding charges 
on
the Fe sites in phases $A_{I}$ and $I_{aa}.$}
\label{Figure6}
\end{figure}

Finally we consider the effect of the lattice distortion of the rung 
with
hopping $t_{a}(1\pm \delta )$ alternating along the $c$-direction and 
we
introduce an elastic energy term $\frac{1}{2}B\delta ^{2}$ per rung. For an 
homogeneous
magnetic state the model reduces to the simple Peierls model considered 
by
Latg\'{e} and Continentino\cite{latg�} and is unlikely to reproduce the
experimental behaviour for reasonable values of $t$ even in the 
undimerized
state as shown in Fig.\ref{Figure4}. As discussed above, the zig-zag 
$I_{a}$
phase strongly favors the related rung distortion and, as expected, it
occupies an important part of the phase diagram as shown in 
Fig.\ref{Figure5}
for a value $B/t_{c}=6$. 
Here we do not consider the more complicated $I_{b}$ phase appearing at
lower $J_{c}$. Phase $I_{a}$ shows two distinct regions, an undistorted 
one
with $\theta _{1}=\theta _{2}$ and a wide distorted one. A phase 
$I_{aa}$
with fully dimerized hoppings ($\theta _{1}=0$, $\theta _{2}=\pi $), 
one
ferromagnetic and one antiferromagnetic bond in each rung, is now 
stabilized
by the distortion (Fig.\ref{Figure5}, below the dashed line). The 
distortion 
$\delta $ for the $F$, $A_{I}$ and $I_{aa}$ phases is shown in Fig.\ref
{Figure6}-a as function of $B/t_{c}$. The existence of hopping 
distortion $%
\delta \neq 0$ in $A_{I}$ requires small values of the elastic term $B/t_{c}\lesssim \sqrt{2}t$. This is easily obtained from the total energy per rung $E$ which reduces to $%
\frac{E}{t_{c}}=-t\sqrt{2(1+\delta ^{2})}$ $+\frac{1}{2}\frac{B}{t_{c}}%
\delta ^{2}$ in the 3-site problem. The $I_{aa}$ phase presents the largest distortion 
among
these phases, showing clearly the bond order related to the 
ferromagnetic
character of the bonds. The corresponding charges on the Fe sites in 
phases $%
A_{I}$ and $I_{aa}$ are shown on Fig. \ref{Figure6}-b. 
We see that the phase $A_{I}$ represents better than others the 
experimental
charges both above ($\delta =0$) and below ($\delta \neq 0$) the 
structural
transition $T_{c}$ i.e $n_{2}$ remains constant equal to $1/2$, while 
$%
n_{1}=n_{3}=1/4$ in the undistorted phase and $n_{3}$ approaches $1/2$
whereas $n_{1}$ decreases in the distorted phase. In the $I_{aa}$ phase 
it
is site-3 which has the largest electronic charge $n_{3}=0.5$ contrary 
to
experimental estimate both above $T_{c}$ and below for $74K<T<T_{c}$.

Our results are consistent with the existence of a $A$-type phase as
proposed at $82K$\cite{bordet} but imply that it persists above 
$T_{c}$. On
the other hand, the $I$-type structure proposed at $10K$\cite{bordet} 
should
present charge ordering and lattice distortion, in contradiction with 
the
recent M\"{o}ssbauer results of Larrea et al.\cite{larrea04}. We have 
shown
that simultaneous spin and charge ordering in qualitative agreement 
with the
experimental behaviour for $T>74K$ occurs from the competition between 
DE\
and SE interactions. The bonding is strongly reinforced by the 
ferromagnetic
correlations, therefore this may induce a lattice instability as 
observed.
Below $74K$, the experimental results\cite{larrea04,bordet,douvalis02} 
are
contradicting and further experiments are required to clarify the low
temperature situation. Our approach has emphasized the importance of 
the
magnetic structure and bring to light the interplay between spin 
ordering,
charge ordering and lattice distortion.

We are grateful to J.\ Dumas, M.\ A.\ Continentino and P.\ Bordet for
helpful discussions. E.\ V.\ acknowledge CONACyT for financial support.

\end{document}